\documentclass[final,12pt,dvipsnames]{elsarticle}

\usepackage{microtype}
\usepackage{amssymb}
\usepackage{comment}
\usepackage{xcolor}
\usepackage{caption}
\usepackage{subcaption}
\usepackage{hyperref}
\usepackage{tabularx}
\usepackage{booktabs}
\usepackage{enumitem}

\newcounter{challengecounter}

\makeatletter
\newcommand\researchquestion[1]{%
  \@startsection{paragraph}{4}{\z@}%
  {1.5ex \@plus1ex \@minus.2ex}
  {-1em}
  {\normalfont\normalsize\itshape}
  {#1}
}

\def\ps@pprintTitle{
 \let\@oddhead\@empty
 \let\@evenhead\@empty
 \def\@oddfoot{\footnotesize\itshape
      Preprint \hfill June 3, 2025}%
 \let\@evenfoot\@oddfoot}

\hypersetup{
	pdftoolbar=true,        
	pdfmenubar=true,        
	pdffitwindow=false,     
	pdfstartview={FitH},    
	pdftitle={},    
	pdfauthor={},     
	pdfsubject={},   
	pdfcreator={},   
	pdfproducer={}, 
	pdfkeywords={}, 
	pdfnewwindow=true,      
	colorlinks=true,       
	  linkcolor=Brown,          
	  citecolor=OliveGreen,        
	  filecolor=magenta,      
	  urlcolor=NavyBlue           
}

\makeatother
\journal{Computers in Industry}
\begin{document}
\begin{frontmatter}
\renewcommand{\thefootnote}{\fnsymbol{footnote}}

\title{Process Mining on Distributed Data Sources}
\author[tud]{Maximilian Weisenseel}
\author[cau]{Julia Andersen}
\author[hub]{Samira Akili}
\author[ub]{Christian Imenkamp}
\author[cau]{Hendrik Reiter}
\author[hub]{Christoffer Rubensson}
\author[cau]{Wilhelm Hasselbring}
\author[tuhh,cau]{Olaf Landsiedel}
\author[uu]{Xixi Lu}
\author[hub]{Jan Mendling}
\author[tud]{Florian Tschorsch}
\author[hub]{Matthias Weidlich}
\author[ub]{Agnes Koschmider}

\affiliation[tud]{organization={Dresden University of Technology (TU~Dresden)},
            country={Germany}}
\affiliation[tuhh]{organization={Hamburg University of Technology (TUHH)},
            country={Germany}}
\affiliation[cau]{organization={Kiel University (CAU)},
            country={Germany}}
\affiliation[hub]{organization={Humboldt-Universität zu Berlin (HU Berlin)},
            country={Germany}}
\affiliation[ub]{organization={University of Bayreuth},
            country={Germany}}
\affiliation[uu]{organization={Utrecht University (UU)},
            country={Netherlands}}

\begin{abstract}
Major domains such as logistics, healthcare, and smart cities increasingly rely on
sensor technologies and distributed infrastructures to monitor complex
processes in real time. These developments are transforming the data
landscape—from discrete, structured records stored in centralized systems to
continuous, fine-grained, and heterogeneous event streams collected across
distributed environments. As a result, traditional process mining techniques,
which assume centralized event logs from enterprise systems, are no longer
sufficient. In this paper, we discuss the conceptual and methodological
foundations for this emerging field. We identify three key shifts: from
offline to online analysis, from centralized to distributed computing, and
from event logs to sensor data. These shifts challenge traditional
assumptions about process data and call for new approaches that integrate
infrastructure, data, and user perspectives. To this end, we define a
research agenda that addresses six interconnected fields, each spanning
multiple system dimensions. We advocate a principled
methodology grounded in algorithm engineering, combining formal modeling with
empirical evaluation. This approach enables the development of scalable,
privacy-aware, and user-centric process mining techniques suitable for
distributed environments. Our synthesis provides a roadmap for advancing
process mining beyond its classical setting, toward a more responsive and
decentralized paradigm of process intelligence.
\end{abstract}






\end{frontmatter}

\section{Introduction}
Process mining is a research domain that focuses on developing and evaluating techniques for extracting insights from event data recorded during the execution of business processes~\cite{DBLP:books/sp/Aalst16}. Examples of such techniques include algorithms for the automated discovery of
process models~\cite{augusto2018automated}, for checking conformance between specifications and recorded
events~\cite{DBLP:books/sp/CarmonaDSW18}, and for predictive
analytics~\cite{DBLP:journals/tkdd/TeinemaaDRM19}.
The obtained operational insights facilitate effective decision making for business process management~\cite{reinkemeyer2020process,badakhshan2022creating}.

Traditionally, process mining assumes that event
data are extracted from information systems, such as enterprise resource
planning systems, that are backed by well-structured relational databases.
These information systems are typically centrally administered within a
single organization. Moreover, they typically log major business activities
and status transitions of process executions, making them well-suited for
process analysis.

For such analysis scenarios, there has been an impressive uptake of process
mining tools in recent years~\cite{grisold2021adoption}.
However, this classical, centralized approach, with the explicit assumption of a single event log as input, offers only a limited view of the full operational picture. It often overlooks the fine-grained behaviors that occur between major process events or outside the scope of traditional enterprise systems.

The adoption of sensor technologies and the rise of the
Internet-of-Things (IoT) are changing this landscape fundamentally. They
enable continuous, real-time observations and monitoring of operational
processes, increasing the transparency of processes from major
status updates to fine-granular
behavior, from structured data to semi-structured or unstructured data, and
from offline, batch-oriented analysis to online, real-time analysis. These
developments open the door to a more fine-granular and dynamic form of
process mining, one that can respond to ongoing events in distributed
environments.
%

Corresponding IoT applications have emerged in various domains. For example, in logistics, services such as
Fleetmon\footnote{https://www.fleetmon.com/services/live-tracking/fleetmon-explorer/}
 use sea ship transponder data for monitoring the movement, loading, and
unloading of vessels.
In healthcare, hospitals 
install Real-Time Locating Systems to track events relating to staff, patients, and
equipment in clinical pathways~\cite{DBLP:journals/is/SenderovichWYGM16}.
Smart city initiatives, such as the ones in Dublin or Warsaw, track information
about traffic events and density of public
transportation~\cite{DBLP:conf/edbt/ArtikisWSBLPBMKMGMGK14}.

The mentioned scenarios have three features in common: (1) they support \textit{complex processes} in a \textit{distributed environment}, meaning they operate within an infrastructure, where process data is collected from geographically or logically distributed data sources; (2) they generate fine-grained, uncertain data, often requiring aggregation and interpretation before becoming meaningful at the process level; and (3) they require real-time user support, such as intuitive visualizations and actionable insights, to guide timely operational decisions.

This means that the sensor data in the above scenarios also integrates into larger business processes. For instance, the event that a sea ship docks in Hamburg is significant for the transportation of hundreds of containers, which require customs handling and distribution in the hinterland, thereby affecting production processes in production plants that rely on raw materials. Similarly, events in a hospital or a delay in urban transportation must be put into context of the underlying processes and the effective decision making of doctors and traffic participants.

This new class of use cases
provides an
opportunity for process mining, but also new technical and conceptual requirements, as follows.
\begin{description}
    \item[Infrastructure-awareness:] Sensor-based systems are physically distributed and, as such, provide an inherently distributed environment for data processing. Here, data aggregation requires near real-time capabilities and resource efficiency for many scenarios.
    \item[Data-awareness:] The traditional assumptions on data in process mining, i.e., an event trustfully representing single activity executions that are ordered and grouped per instance of a process, no longer hold. Sensor data requires aggregation and interpretation over several levels of abstraction, such that eventually meaningful events are obtained that are relevant at the process level.
    \item[User-awareness:] To effectively support process-related decision making based on sensed event data, new representational capabilities are required. Put differently, classic models for dependencies that focus on isolated process instances are not suited to guide and inform users in their decision making. Solutions, therefore, need to facilitate traceability of the analysis along with effective representations and visualizations.
\end{description}

In this paper, we introduce the concept of \emph{process mining on distributed data sources}
as our main contribution to address the above emerging challenges.
We lay the foundations for this
new class of process analysis techniques, designed to operate effectively
across fragmented, heterogeneous, and dynamic data landscapes. We envision techniques for decentralized, online process mining
that process the data as close to the data sources as possible
and that work on partial data, while allowing timely and continuous analysis.
The data analysis will be enforced in the light of the data's utility,
preserving the accuracy and completeness of process mining results as much as possible
by embedding the uncertainty of data in the entire process mining pipeline,
making data quality issues explicit in the analysis and protecting the data through provable guarantees.

We argue that advancing research in the area of process mining on
distributed data sources requires establishing new methodological
foundations.
A central characteristic of this setting is that the infrastructure, data,
and user perspectives are deeply interwoven and cannot be treated in
isolation. This
interdependence not only introduces complexity but also creates opportunities for collaborative solutions. For example, abstracting data at the source can reduce the risk of disclosing personal information, but may also increase uncertainty in the resulting analysis. Similarly, performing online analysis close to the data source enables timely insights, yet introduces trade-offs between responsiveness and potential privacy breaches.
To achieve this, future research must integrate expertise from
process management, data and software engineering, distributed systems, and privacy-preserving computation. Addressing challenges at the intersection of these domains offers the potential for substantial research impact by laying the groundwork for the next generation of process mining techniques.

The remainder is structured as follows.
In Section~\ref{sec:2}, we
describe business processes from different industries that already utilize sensors today. We use these use cases to outline the opportunities and
requirements for process mining on distributed data sources. Based thereon,
in Section~\ref{sec:researchfields}, we discuss the research fields related
to these requirements, along with research questions that need to be
addressed. Linking these questions to the state of the art in the field, we
identify conceptual and technical research gaps. Section~\ref{sec:4}
then synthesizes specific directions for future research to address these
gaps, before Section~\ref{sec:5} concludes the paper.

\section{Business Processes and Distributed Data Sources}\label{sec:2}
Business processes increasingly involve distributed data sources spanning
organizational, technical, and geographical boundaries. These sources enable
more informed, timely, and context-aware decision-making. In this section, we
highlight three domains---logistics, healthcare, and smart manufacturing---to
illustrate how distributed data enhances both operational control and
strategic planning through process mining.

\subsection{Logistics}
Business processes in logistics increasingly make use of advanced real-time
decision making. A real-world transport scenario is described
in~\cite{di2016detecting}. It refers to a multi-modal transportation of
high-value technical components that are shipped by airplane, then picked up by
trucks, and delivered to a production plant. As a first step, the
technical components are transported to John F. Kennedy International airport
(New York, USA). There, they are loaded onto a freight airplane. This airplane
flies to Schiphol Airport (Amsterdam, The Netherlands). Trucks wait at the
airport for the arrival of the flight to pick up the delivery. These trucks are
operated by a logistics service provider. They
transport the goods to the destination plant in Utrecht.

In general, this process operates smoothly with predictable arrival times of the flight. Difficult are instantaneous changes of the weather conditions
\cite{di2016detecting}. In rare cases such as thunderstorms, Schiphol Airport
must be closed for arriving flights. These are typically rerouted to Brussels. A
challenge is the fact that neither truck operator nor the customer have a direct
channel of communication with the flight operator. If the customer knew right on
time, they could cancel the trucks in Amsterdam and order trucks in Brussels.

The recent availability of real-time flight trajectory data bears the potential
to predict the rerouting of flights based on the continuous observation of
positions. To that end, real-time flight data has to be immediately processed. Di
Ciccio et al.\ propose a support vector machine tailored to this
problem~\cite{di2016detecting}. It takes as input the stream of position events
over a predefined time window and calculates unusual bearings of the flight. It
is a requirement that this data has to be processed at real time. The analysis
must also reflect uncertainty of position information, which is triangulated from
noisy transponder readings of a distributed network of antennas.

\subsection{Healthcare}
In healthcare, clinical pathways describe how patients are treated. For instance,
a cancer treatment process at an outpatient clinic involves several steps, such as
patient registration, a blood draw, a vital exam, an infusion, and the
release~\cite{DBLP:journals/is/SenderovichWYGM16}. These activities involve
different types of medical staff and equipment. For instance, infusions need to be
prepared at a local pharmacy per patient, and are administered by dedicated
infusion nurses in specific rooms that are equipped with infusion chairs.
Moreover, the treatment of a patient is scheduled, i.e., appointments for the
separate activities are booked in advance. In practice, however, there are many
reasons for dynamic changes to the schedule, from patient no-shows, through
unexpected outcomes at vital exams and non-availability of clinical experts, to
operational delays caused by the pharmacy.

For this process, operational management concerns different time horizons. On the
one hand, dynamic adaptation of the patient schedules during a single day is
required to react to the aforementioned situations, striving for a high quality of
service as well as optimal resource utilization. On the other hand, decision
making relates to the overall efficiency of the process from a long-term
perspective, i.e., considering overbooking of treatment slots or stand-by
personnel.

In order to support operational management, in this scenario, the data stored in
various information
systems may be leveraged, e.g., an appointment booking system, a staff planning
system, and a system managing electronic health records of patients. However, data
from these systems is typically only available at a coarse-grained level (e.g.,
recording that an infusion has been done, not providing progress information),
often suffers from data quality issues (e.g., manually inserted timestamps of
treatments turn out to be relatively unreliable), and lags behind the actual
process execution (e.g., data on treatments is inserted by medical staff only at
the end of their shift). In the case described
in~\cite{DBLP:journals/is/SenderovichWYGM16}, therefore, the clinic also relies on
data from a real-time locating system (RTLS) that tracks the locations of
transponders that are worn by patients and medical staff, and attached to medical
equipment, with a resolution of around three seconds. Such sensed data opens a
completely new perspective for process analysis, in terms of immediate,
fine-granular control. Yet, leveraging the RTLS data also imposes new challenges,
as the data first needs to be abstract to be interpreted in terms of process
progress~\cite{DBLP:conf/caise/SenderovichRGMM16}, which is an inherently
uncertain procedure that also needs to be realized in a privacy preserving manner.
Moreover, from a technical point of view, data analysis shall rely on online
algorithms to facilitate immediate operational control, while the data is emitted
by a distributed sensing infrastructure, with RTLS receivers being hierarchically
organized according to the building structure.

\subsection{Smart Manufacturing or Factory}
In smart manufacturing or factory immediate decision-making is necessary in order to detect anomalies or to make predictions. The GlassFactory case study refers to an Industry 4.0 glass manufacturing company~\cite{DBLP:journals/sosym/MontiMLKM24}. It involves several actors (or objects respectively): raw material batches, a warehouse, heating, processing, cooling machines, a converter machine, two mobile robots for transport, and ERP/MES systems for managing business and production processes. The glass production process follows a strict sequence: heating, shaping, and cooling, or alternatively, a single-step transformation using the converter. All machines and robots are sensor-equipped and API-accessible, allowing real-time monitoring and event execution based on specified requirements. The challenge of this scenario is to coordinate multiple objects in real-time while integrating heterogeneous data from sensor-based devices and an ERP system. So we observe a distribution of operations with tasks. However, an efficient coordination of the process requires an understanding of the process model (i.e., which activities depend on which activities or what common data does the execution of activities require?).

A data-oriented approach might bridge the gap between the infrastructure, data, and user layer in this scenario. The data layer supports efficient filtering of events at the infrastructure level. At the same time, the user level visualizes the dependency of the data and also allows for better event coordination.



\subsection{Common Novelties for Process Mining on Distributed Data Sources}
\begin{table}[bt]
\caption{Summary of use cases involving process mining on distributed data sources}
\label{tab:usecases}
\scriptsize
\newcolumntype{R}{>{\raggedright\arraybackslash}X}
\begin{tabularx}{\textwidth}{p{21mm}RRR}
\toprule
\textbf{Domain} & \textbf{Characteristics} & \textbf{Data Sources} & \textbf{Decision Type} \\
\midrule
Logistics &
  Multi-modal transport, weather-sensitive, inter-organizational &
  Flight trajectories, weather data, airport and logistics systems &
  Real-time rerouting and coordination \\
\midrule
Healthcare &
  Multi-step clinical pathways, patient-specific, resource-constrained &
  RTLS, EHR, scheduling and staff systems &
  Dynamic rescheduling and long-term planning \\
\midrule
Smart Manufacturing &
  Distributed assembly lines, sensor-driven, step dependencies &
  Sensors, video cameras, MES/SCADA systems &
  Online monitoring and coordination \\
\bottomrule
\end{tabularx}
\end{table}
Table~\ref{tab:usecases} summarizes the three use cases
by outlining their process characteristics, data sources,
and the decision-making capabilities supported
through process mining on distributed data sources.

What is novel across these scenarios is the use of process mining on
distributed data sources to go \emph{beyond retrospective analysis}. These
approaches support dynamic monitoring, real-time decision-making, and
continuous process optimization. This represents a significant step beyond
traditional, centralized analytics and unlocks new possibilities for
process-aware, data-driven decision support.



\section{Process Mining on Distributed Data Sources: Research Fields and Approaches}
\label{sec:researchfields}
A fundamental assumption in traditional process mining is the existence of a central event log,
where event data is available as a single file---static, totally ordered, and accurate~\cite{DBLP:books/sp/CarmonaDSW18}.
This idealized view, however, diverges from reality of distributed, dynamic, unordered, and uncertain sensor data. Despite this reality, process mining has predominantly focused on \emph{how} to mine efficiently. Recognizing that creating an event log involves data transfer, ordering, and cleaning, it is essential to shift the focus from \emph{how} to mine efficiently to \emph{where} to mine efficiently.

Process mining on distributed data sources starts on the observed system itself. By shifting process mining closer to the data sources,
we aim to achieve strong privacy, an extensible and scalable infrastructure, and online analysis. Privacy concerns arise from centralized data storage, which invites misuse and breaches.
Realizing process mining on distributed sources minimizes data misuse by ensuring data processing is purpose-limited and minimal. Scalability is challenged by the ever-increasing volume of data, where centralized data collection is inefficient.
Process mining on distributed data sources starts processing events right at the data sources. It is, hence, more efficient,
reduces bandwidth and offers better scalability and fault tolerance.
While online processing is essential for providing immediate responses, most algorithms are not designed for this purpose. In contrast, process mining on distributed sources enables online analysis directly at the data source.


\begin{figure}[t]
  \centering
  \includegraphics[width=\textwidth]{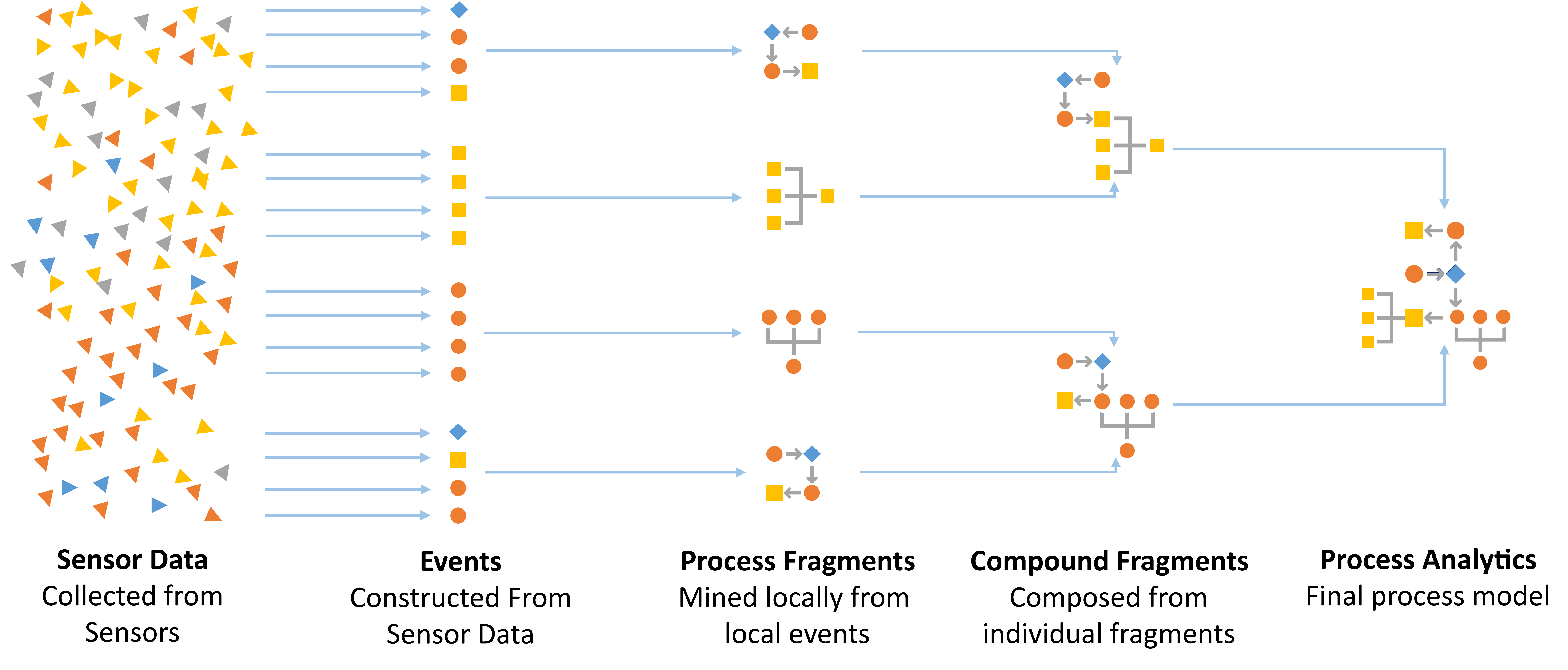}

  \caption{Process mining on distributed data sources.}
  \label{fig:setting}
\end{figure}

Process mining on distributed data sources introduces a range of novel research fields while challenging established ones. The following section provides an overview of these
research fields and outlines corresponding technical
approaches. The research fields can be grouped in several core themes:
processing near the edge to deal with data velocity and privacy,
ensuring scalability across heterogeneous infrastructures,
abstracting low-level sensed data into process-aware insights
embedding privacy directly into collection and processing pipelines
supporting explainability of data handling and outcomes,
and developing effective visual representations for complex, distributed processes.

Each subsection introduces associated research questions intended to guide the
development of new process mining techniques suitable for distributed
environments. An overview of the main research fields and associated research
questions is provided in Table~\ref{tab:researchfields-overview}.

\begin{table}[p]
\centering
\caption{Research fields and guiding research questions in process mining on distributed data sources}
\label{tab:researchfields-overview}
\newcolumntype{R}{>{\raggedright\arraybackslash}X}
\scriptsize
\begin{tabularx}{\textwidth}{lR}
\toprule
\textbf{Edge Processing} & Bringing process mining closer to data sources enables real-time analysis and reduces reliance on central infrastructure. Local mining on IoT and edge devices addresses challenges such as data velocity.
    \begin{itemize}[left=0em, itemsep=-0.5em, topsep=0.5em]
      \item How can we achieve locality-aware process mining?
      \item How can we achieve resource-efficient process mining?
      \item How can we achieve adaptive process mining in dynamic environments?
    \end{itemize}\\
\midrule
\textbf{Scalability} & Process mining on distributed sources must scale to handle high-volume, high-velocity event streams across heterogeneous infrastructures. Efficient distributed processing is key to maintaining responsiveness in online and streaming scenarios.
\begin{itemize}[left=0em, itemsep=-0.5em, topsep=0.5em]
  \item What factors influence scalability in distributed settings?
  \item Which techniques improve scalability across infrastructures?
\end{itemize}\\
\midrule
\textbf{Abstraction} & Process mining on distributed data sources often begins with low-level sensor data that must be abstracted into meaningful activity-level information. Effective abstraction techniques are needed to bridge the semantic gap between raw data and process models, while supporting privacy and decentralization.
\begin{itemize}[left=0em, itemsep=-0.5em, topsep=0.5em]
  \item How does data abstraction impact participant privacy?
  \item How can abstraction and exchange be balanced across sources?
\end{itemize}\\
\midrule
\textbf{Privacy} & Unlike traditional settings, process mining on distributed sources must protect sensitive data not only at publication but also during collection and runtime. Privacy-aware mining techniques must consider local data access, adversary models, and privacy metrics suited for distributed architectures.
\begin{itemize}[left=0em, itemsep=-0.5em, topsep=0.5em]
  \item What are suitable adversary models and threats?
  \item What privacy metrics suit process mining on distributed data sources?
  \item Which techniques preserve privacy without full case visibility?
\end{itemize}\\
\midrule
\textbf{Explainability} & The integration of noisy or heterogeneous sensed data in process mining on distributed data sources complicates the interpretation of results. Transparent outlier handling and traceability across the pipeline are needed to explain and trust process mining outcomes.
\begin{itemize}[left=0em, itemsep=-0.5em, topsep=0.5em]
  \item How to propagate outlier handling throughout the pipeline?
  \item How to use distribution to improve outlier detection?
\end{itemize}\\
\midrule
\textbf{Visualization} & Visualizations for process mining on distributed sources must accommodate heterogeneous, large-scale data while supporting time, location, and multi-entity perspectives. Effective layouts and interactions are essential for analysis.
\begin{itemize}[left=0em, itemsep=-0.5em, topsep=0.5em]
  \item What structures and views support distributed analysis?
  \item How to design layouts that scale with data variety and volume?
\end{itemize}\\
\bottomrule
\end{tabularx}
\end{table}

\subsection{Edge Processing}


Established approaches to process mining, commonly, assume that data is readily available for mining at a central location with vast storage and computing capabilities \cite{DBLP:books/sp/Aalst16}.
This assumption holds less and less given following trends:
(1)~Data is increasingly sourced directly from numerous geographically distributed IoT devices such as sensor nodes.
Thus, IoT devices are becoming common sources for process mining in, for example, digital factories, healthcare, and smart cities, and complement databases and ERP systems that are the default data sources today \cite{DBLP:journals/corr/abs-1811-00652, 10.1145/3410566.3410605, DBLP:conf/caise/SenderovichRGMM16, Sztyler2016, 7549355}.
(2)~Already nowadays, data velocity is a challenge and many IoT devices generate significantly more data than they can store or transfer to the cloud\footnote{IDC White Paper – https://www.seagate.com/www-content/our-story/trends/files/idc-seagate-dataage-whitepaper.pdf}.
(3)~Privacy considerations often prevent data collection at a central location.
Overall, the geographically distributed nature of IoT sensors, their high data velocity and often privacy considerations make it practically impossible to collect large-scale process data at a central location for mining.

To address the above challenges, we argue that it is essential to push process mining as close to the data sources as possible by distributing it over geographically scattered IoT sensor nodes, edge and cloud devices, enabling scalable and privacy-aware process mining.
To achieve this goal, it's essential to devise new approaches to (1) locality-aware process mining on distributed data sources and (2) process mining on resource-constrained IoT devices.

In the following, we briefly introduce our approach to process mining on distributed data sources.
In a distributed IoT setting, each sensor node will locally collect an event log and compute a partial mining result, i.e., process fragments, based on its local log, see Figure~\ref{fig:setting}.
Such a fragment is an aggregate of events observed by the sensor and their relationship to events shared by nearby sensors, matched, for example, via event IDs.
Each fragment represents a partial data-flow graph of the process, which the sensor then forwards to neighboring sensors or edge devices, where it will be merged with process fragments of other sensors in a stepwise manner.
With each step in this analysis chain, more fragments will be added in a distributed manner, forming compound fragments.
The partial process graph will grow step by step to become a complete process model.
Finally, at an edge or cloud device, the complete process graph will be available for analytics including, for example, process discovery and conformance checking.
Overall, the result is a distributed data-flow graph, and we map its computation onto physically distributed and heterogeneous IoT devices. We capture this potential by means of the following research questions:

\researchquestion{How can we achieve locality-aware process mining on distributed data sources?}
A first challenge is to devise locality-aware process-mining algorithms in which the data-flow aligns efficiently with the underlying topology of a physically distributed network of, for example, sensor nodes. 
For this, we need to devise new approaches to locality-aware process mining on geographically spread out sensor nodes.
This awareness of the locality of data and data sources in a distributed mining setting is a novelty when compared to state-of-the-art process-mining algorithms, which commonly assume that data is readily available at one central location.
They are not designed to operate on geographically distributed data sources, such as a network of IoT sensors.


\researchquestion{How can we achieve resource-efficient process mining?}
A second challenge is to devise resource-efficient process-mining algorithms that match the limited, often heterogeneous computing-resources in a network of physically distributed IoT sensors, edge and cloud devices.
This is essential as today's process mining algorithms are commonly computationally heavyweight and their resource-requirements are a stark contrast to such resource-constraints.
The challenge is not only to split the process-mining algorithms into distributed variants but also to split these hierarchically where, for example, simpler tasks are mapped to the less powerful devices and complex ones, i.e., with high memory or computation requirements, are either split-up and distributed or---if such a splitting-up is not feasible---placed on more powerful edge and cloud devices.

\researchquestion{How can we achieve adaptive process mining in dynamic environments?}
Our final challenge in this direction is to automate the distribution and mapping of the algorithms devised in this project onto dynamic IoT networks and to dynamically adapt these to changing requirements, resources, and network topologies.
Complex, real-world processes, such as a production line in a smart factory, are constantly evolving.
For example, a production line gets extended due to high demand, changed based upon customer request or parts get replaced due to wear and tear. As a result, both the process we are mining and the IoT devices, which are sensing it, change.
To address this challenge, we need to devise new approaches to automatically adapt our algorithms for process mining on distributed data sources to such dynamics.
Thus, we need to dynamically adapt and reconfigure both the placement of mining functionality and the data-flow graph of process fragments and compounds.

\subsection{Scalability}
Scalability is a central aspect of process mining~\cite{leemans_scalable_2018}, i.e., dealing with volume, velocity, and variability of input data, especially in online settings using event streams, with real-time processing requirements. In this context, the events should be processed immediately as they arrive in a continuous stream.
Process mining should not be restricted to offline analysis, it should also provide online operational support~\cite{van2011process}.
Thus, scalable processing of continuous event streams and process fragments on cloud infrastructures is required for efficient and effective stream process mining.

Scalability refers to the ability of a software system to sustain increasing workloads with adequate performance provided that hardware resources are added~\cite{brataas_scalability_2017,ESE2022}.
When considering continuous streams of events, often an integration of multiple such streams is required~\cite{Henning2020,JSS2024}. With the traditional approach to process mining of first writing all events to a (relational) database and then querying this integrated database, processing the events is straightforward. Processing continuous event streams raises new challenges. This is in particular the case when requirements for scalability have to be considered.
An approach could be hierarchical data aggregation that incrementally combines partial analysis results obtained for process fragments, processed from edge devices to cloud-based frameworks. Scalability is achieved by incorporating the resource capabilities of these layers in deployment decisions. We capture this potential by means of the following research questions.

\researchquestion{Which factors influence scalability?}

The scalability of systems depends on the specific environment in which a system is deployed. For example, cloud environments offer faster access to high-performance computing resources compared to data processing systems operating at the edge. Edge computing presents unique challenges due to the frequent network communication requirements of edge devices. Additionally, these devices often exhibit lower processing power and increased susceptibility to failures. Consequently, chaos engineering plays a pivotal role in assessing scalability. Another layer influencing scalability is the nature of the data that is being processed. For example, processing images from which activities must first be extracted is computationally more intensive than processing activities directly extracted from an information system. However, the variability and temporal characteristics of the data can also affect the processing efficiency.

\researchquestion{Which techniques make process mining on distributed sources more scalable?}

To effectively manage the increasing volume of data, novel techniques are required to process this data efficiently. Potential techniques include parallelization, sampling, and batching. To support parallel processing, new methods must be devised to merge partial results consisting of Petri nets or process trees. However, these methods come with various trade-offs. Sampling can lead to inaccurate process models, batching introduces higher latency, and parallelization increases resource consumption. It is essential to quantify these trade-offs to enable informed decision-making based on specific application requirements.

\subsection{Abstraction}
In sensed event data, events no longer denote explicit activity executions as
part of a particular instance of a process, as commonly assumed in process
mining. Rather, they denote
low-level information such as whereabouts of persons or properties of
resources. As such, there is a need for data abstraction that aggregates
low-level events and interprets them in terms of the process to analyze. At the
same time, shifting to distributed data sources, data is no longer
available at a single location, but needs to be exchanged explicitly between
the involved entities.

In recent years, a wide range of techniques to abstract event data have been
proposed, as reviewed in~\cite{DBLP:journals/widm/DibaBWW20}. Specifically,
existing approaches exploit generic data mining methods, such as clustering
(e.g.,~\cite{gunther2006mining}) and supervised learning
(e.g.,~\cite{tax2016event}), or incorporate richer
information, reaching from fine-granular behavioral
patterns~\cite{DBLP:journals/is/MannhardtLRAT18} through process
models~\cite{DBLP:journals/sosym/BaierCMW18}, to comprehensive domain
knowledge~\cite{DBLP:conf/caise/SenderovichRGMM16}.
Moreover, approaches for supervised learning of queries over sequences of
events~\cite{DBLP:conf/debs/MargaraCT14,DBLP:journals/pvldb/GeorgeCW16,disces},
may
be used to formulate event abstractions for process mining. However, we note
that all these approaches assume that the event data is available at a single,
centralized location. Any distribution of the sources of the respective event
data is neglected.

Moreover, sensed event data can be expected to contain information that is
sensitive from a privacy point of view. For instance, while there may be good
reasons to track the execution of a treatment step by some medical staff in a
hospital, deriving this information from the whereabouts of the medical staff
(as detailed above and in~\cite{DBLP:journals/is/SenderovichWYGM16}), arguably
requires careful and responsible data management. As such, the need for the
abstraction of event data from distributed sources induces not only conceptual
challenges, but also provides an opportunity to achieve privacy-aware process
mining. That is, recent advancements on event data sanitization in process
mining~\cite{DBLP:journals/bise/MannhardtKBWM19,
DBLP:conf/bpm/Fahrenkrog-Petersen20,
DBLP:journals/dke/RafieiA21,DBLP:journals/dke/FahrenkrogPetersenAW23}, which
all work on traditional notions of event data that are available at a single
location, may be complemented with techniques that explore the interplay of
privacy guarantees and distributed abstraction of event data. We capture this
potential by means of the following research questions.

\researchquestion{How to achieve privacy by constraining the abstraction of
event data for process mining?} This question aims at an understanding of how
the information loss induced by common techniques to
aggregate and interpret sensed event data influences the privacy of process
participants. To this end, we propose consider two directions to achieve
privacy guarantees: (i) by transforming
the data, before or after some abstraction function is or has been applied; or
(ii)~by changing the abstraction function. The former may rely on noise
insertion into aggregated intermediate data representations, as well as
grouping and generalization of events. The latter, in turn, may comprises
changes to the event patterns used in data abstraction as well as stochastic
approaches to map the found patterns to activity executions. Either way, data
utility shall be preserved as much as possible, which suggests formulating the
privacy-aware abstraction of event data as an optimization problem.

\researchquestion{How to achieve privacy by constraining the exchange of event
data across distributed sources?} This question strives for an
understanding of the
interplay of event abstraction (e.g., in terms of the resulting granularity of
the event data) and properties of data exchange schemes (i.e., what data is
shared with whom and when?). Here, our vision is that the privacy-aware
abstraction as mentioned above is grounded in a networked setting, by assigning
abstraction steps to data sources. Then, the optimization of the utility of the
abstracted data, while satisfying privacy guarantees, may incorporate
constraints that limit the data exchange.

\subsection{Privacy}
Privacy is an established topic in the process mining community.
However, the current approaches are largely influenced by a data mining perspective.
As a result, privacy concerns are often framed in terms of the risk of identifying individuals in published event logs~\cite{Majid2021,nunezvonvoigt20reidentification} or models~\cite{Maatouk2022}.
Expanding the perspective to include the process of data collection reveals additional privacy challenges:
Relying on centralized storage of event data can increase the risk of unintended data use or unauthorized access.
In this setting, everyone with access to the data can analyze it for any purpose without overcoming any additional safeguards.
Furthermore, it is an enticing target for attackers.
Process mining on distributed data sources offers an alternative that enables
privacy mechanisms to be integrated directly at the point of data collection.
By processing data close to the source and restricting data sharing to what
is strictly necessary for a given purpose, this approach naturally supports
principles such as data minimization and purpose limitation.
Consequently, we want to move from privacy protection at the point of publication to privacy protection at the point of collection.

\researchquestion{What are relevant adversary models and respective privacy threats when process mining on distributed data sources?}
The main goal of an adversary is to observe the whole case of a specific user, so the adversary learns all activities of a user and in which order they have been executed.

We define two adversaries for process mining on distributed data sources.
The first adversary (\emph{public data adversary})
targets published data, e.g., intentionally shared data, which has already been processed and might be anonymized \cite{Majid2021,nunezvonvoigt20reidentification, Maatouk2022}.
Here the observations are the published event logs.
The second adversary (\emph{private data adversary})
focuses on the collected but unpublished data. Here, the adversary controls a subset of the data sources or nodes in the network, and the observations are the information observed at the corrupted nodes. Potential adversaries could be internal data scientists misusing the data (for example, for a different purpose than the one it was collected for) or an adversary with access to a subset of the nodes (e.g., a manufacturer or an administrator).

\researchquestion{What privacy principles and techniques enable process mining in distributed settings where no entity observes the entire case?}

General principles like filtering, aggregation, separation, and anonymization \cite{hoepman2014privacy}
appear promising for enabling privacy-aware process mining in distributed settings, where no single entity observes the entirety of a case. Filtering limits the transmitted data
by forwarding only common events or typical arcs. Additionally, filtering protects outliers and enhances process model utility
by focusing on representative patterns.
Similarly, aggregation combines data from multiple users or sensors
into an abstract form that removes user-specific details.
For example, partial footprint matrices computed at different nodes
can be combined to describe the entire process~\cite{van2013decomposing, andersen2024edgeminerdistributedprocessmining},
which reduces the sensitive information exchanged
while preserving process structure.

\researchquestion{What are useful privacy metrics for process mining on distributed data sources?}
An orientation offer the three main privacy metrics uniqueness~\cite{sweeney2000simple, golle2006revisiting}, \emph{k}-anonymity~\cite{sweeney2002k}, and differential privacy~\cite{dwork2006differential} and adjust them for the setting of process mining on distributed data sources. Nu\~nez von Voigt et al.~\cite{nunezvonvoigt20reidentification} analyzed the uniqueness of cases and case attributes in different event logs to determine the reidentification risk. They computed the fraction of unique cases (respectively, case attributes) and compared them to the total number of cases. Nevertheless, uniqueness does not indicate the reidentification risk of non-unique cases. The reidentification risk differs if two cases are identical or if thousands of cases serve as the anonymity set. Therefore, the k-anonymity of traces should be considered in addition. Finally, we propose considering the differential privacy of traces. A differential private formulation of the identification risk could be: The information an adversary can learn from including a single case in the event log should be bounded. Finally, in the setting described above, we measure the privacy threat by comparing the background knowledge to the posterior knowledge. This enables measuring how much the adversary has learned.

\subsection{Explainability}
The integration of sensed data increases the amount of outliers and noise. Such anomalies, in general, hamper the usefulness of analysis results~\cite{Berti2024}. Yet, mechanisms such as differential privacy rely on a perturbation of the data and, hence, may exploit the inherent presence of anomalies~\cite{Mannhardt2019}. Against this background, this project strives for means to detect and quantify anomalies in event data. Based on these measures, models to control the proactive perturbation of data for privacy-awareness based on the extent of anomalies that are present already will be developed. By making these design choices explicit, the explainability of the analysis results is improved~\cite{Berti2024, Koschmider2021a}. We
capture this potential by means of the following research questions.

\researchquestion{How to propagate outlier pre-processing throught out the complete analysis pipeline?}
A time-consuming step in the analytics pipeline from processing (raw) data to discovering knowledge is data pre-processing in terms of e.g., outlier detection~\cite{https://doi.org/10.18420/modellierung2024_012}.  We consider outliers as divergent data that is out of range of (process) behavior that we expect~\cite{Koschmider2021a, Berti2024, VITALE2025112970}. This can, on the one hand, point to insignificant data or, on the other hand, to interesting and useful information about the underlying system. Hence, distinguishing the essence of outliers in terms of undesired or unwanted behavior versus
surprisingly correct and informative data is of particular interest to \textit{explain} the analysis result. To provide a solution, outlier information (i.e., detection, filtering, removal, completion) might be propagated to each step of the analysis (i.e., from pre-processing through abstraction to visualization) in terms of meta-data annotations like confidence scores~\cite{Birihanu2025}. The annotations allow guidance on the impact of the range of outliers on the discovered process model. While meta-data annotation might be one suitable solution, more research effort is needed to come up with additional solutions.

\researchquestion{How to exploit distribution to improve outlier detection?}
Mostly, related works use data from one central source or nodes respectively like an event log in combination with context information to identify outliers~\cite{Berti2024, Koschmider2021a, DBLP:journals/corr/BohmerR17, 10.1007/978-3-642-01862-6_13}. Information from distributed sources is not exploited to improve outlier prevention, mitigation, detection nor repair. By combining information from multiple sources, it is possible to obtain more complete information for classifying outliers as normal or abnormal behavior~\cite{9838396}. The challenge, however, remain to efficiently collect data from distributed nodes, which continuously process a large volume of data. Since nodes are usually equipped with low batches for storing, efficient and scalable pre-processing is hampered.

\subsection{Visualization}
Process mining on distributed sources emphasizes the importance of incorporating contextual factors, such as \emph{time} and \emph{space}, into visual representations to derive meaningful insights from sensor data about their underlying processes (cf.,~\cite{ye2012iotcontextreview,grisold2024contextprocessmining}). In addition, data from distributed sources are oftentimes heterogeneous and high in volume, posing both a technical and a conceptual challenge for effective visual representations. Conventional visualizations in process mining in a centralized setting do not directly apply to a distributed context for two main reasons. First, the existing range of visual representations in process mining is limited to control-flow-centric process models, like Petri Nets and Directly-Follows Graphs~\cite{augusto2018automated,van2011process}. In other words, they do not sufficiently address the needs of process analysts in a distributed setting. Second, current visual representations are also established based on the assumption that processes are structured and partially sequential, consequently struggling to handle event logs with high variation. In light of these challenges, we derive the following research questions.

\researchquestion{What are suitable data structures and corresponding visual representations for supporting users applying process mining for distributed data sources?}
Recent research challenges the typical one-dimensional perspective of a process in various ways. For example, multi-entity event data formats like OCEL~\cite{ocel} and event knowledge graphs~\cite{esser2021multidimensionalgraph} address the issues of convergence and divergence~\cite{van2019object} found in traditional event logs by expanding the number of case notions in the analysis. Multidimensional variant analysis~\cite{nguyen2018multipersvariants,cremerius2023datavariants,adams2022ocelvariants,rubensson2024contextvariants,adams2024supervariants} questions the presumption of control-flow-based process variations and redefines them for multiple dimensions. Also, recent work on process layouts has proposed other types of visual arrangements to incorporate other dimensions, e.g., timeline-based approaches~\cite{suriadi2015waitingtimes,kaur2024timeline} or spatial process arrangements~\cite{de2012visual,corea2024geoprocess}. Therefore, multidimensional analysis in process mining remains underexplored, and corresponding visual representations are limited.

\researchquestion{How can layout strategies be designed to ensure that process-centric visual representations for process mining on distributed sources remain effective with increasing data volume and heterogeneity?}
Visual analytics handles the problem of data volume and heterogeneity by applying various strategies~\cite{du2017volumevarietystrategies}, e.g., by aggregating event patterns~\cite{monroe2013eventflow}, or by using hierarchical clustering rendering techniques~\cite{zinsmaier2012interactiverendering}. Such techniques are rarely applied in a process mining setting, but few exceptions exist for optimizing layouts.
An example of this is the \emph{fuzzy miner}~\cite{gunther2007fuzzyminer}, for which significance and correlation metrics are used to simplify process graphs. Other examples are work that uses layout optimization techniques to structure graphs to enhance readability~\cite{yang2004workflowsugiyama,sonke2018linearlayout,mennens2019processsugiyama}. Still, this area remains a relatively untouched topic of process mining.

\section{Collaborative Research}\label{sec:4}
In the following, we provide a unified perspective on process
mining in distributed environments. The goal is to outline how research on
process mining must evolve to address the combined demands of infrastructure,
data, and user awareness under changing conditions. This synthesis reveals
not only the need for integrated algorithmic and architectural solutions but
also highlights collaborative research opportunities.

These collaborative opportunities become apparent when taking a step back and
considering the broader shifts. Based on the preceding discussions, we
observe several fundamental shifts in how process mining is applied in
distributed environments. While established process mining, is usually retrospective and executed on a central event log,  process mining on distributed data sources is distributed, online and mines on sensor data.
Thereby, process mining on distributed data sources introduces the following three shifts:
\begin{description}
  \item[Offline to online:] Instead of mining events from static records real-time data streams are processed in order to enable real-time monitoring and decision making
  \item[Event logs to sensor data:] The entities do not work on abstract events, instead, they rely on fragmented and heterogeneous data sources
  \item[Centralized to distributed:] The data sources are distributed either over an assembly line, a hospital or the globe and involve interdependent process steps across multiple actors or locations
\end{description}
At the same time, the system model for process
mining on distributed data sources emphasizes that the infrastructure, data,
and user perspectives can no longer be considered in isolation. Instead,
effective process mining solutions must account for all three dimensions in
an integrated manner. Together, these developments fundamentally redefine the
requirements for process mining techniques and induce intertwined research
fields.


\begin{figure}[t]
    \centering
\includegraphics[width=\textwidth]{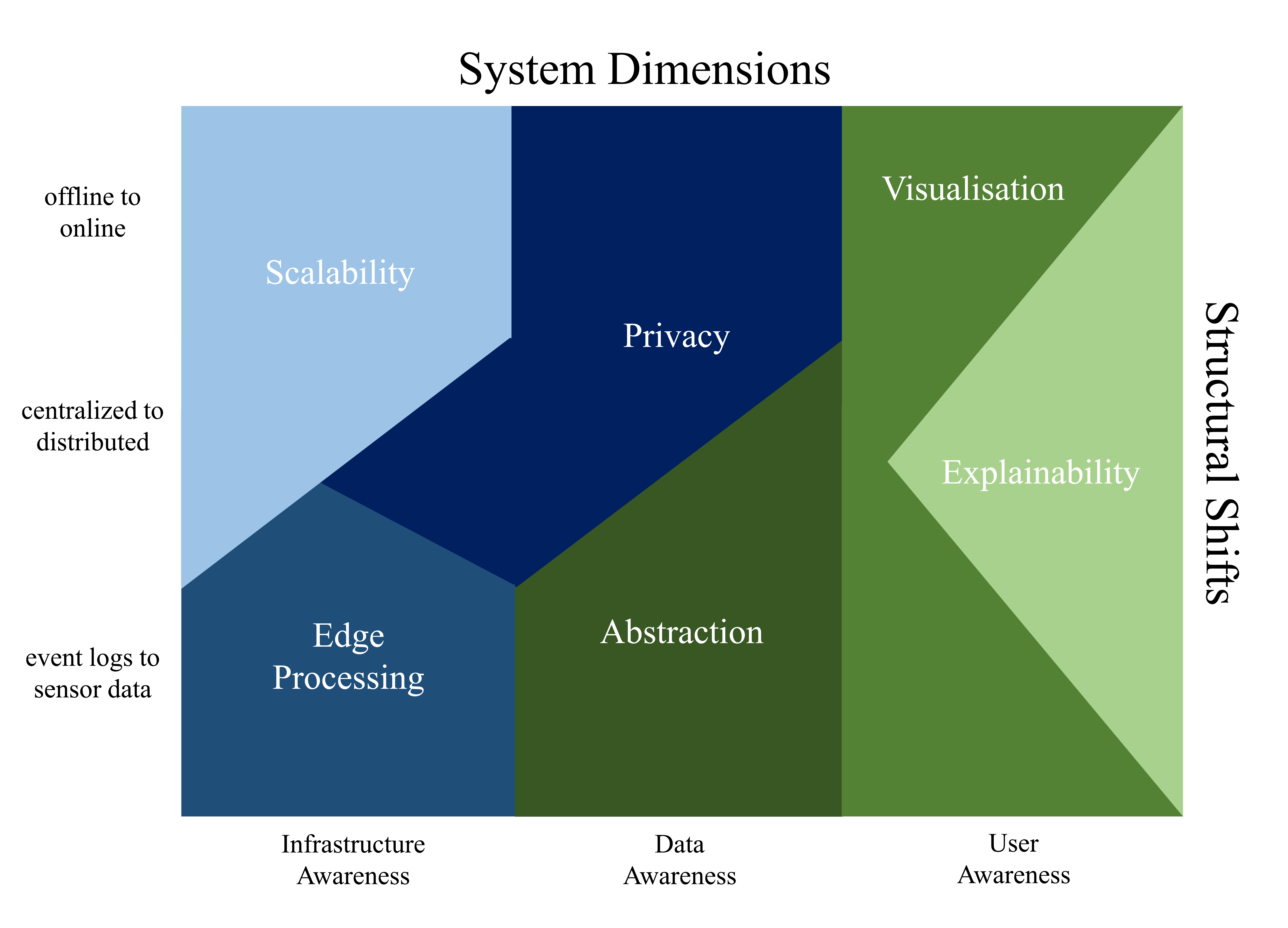}
    \caption{
    Cross-cutting research fields in process mining on distributed sources,
    and induced shifts.}
    \label{fig:matrix}
\end{figure}

Each research field previously discussed---edge processing, scalability, abstraction,
privacy, explainability, and visualization---spans one or more of these
dimensions and often emerge at their intersection. To structure these
research fields and clarify their interrelations, future research can be drawn on a
problem and objective matrix, as illustrated in Figure~\ref{fig:matrix}. It
highlights how different areas of research complement one another through
their respective focus on particular system dimensions, while collectively
advancing the state of process mining.

We recommend that each research field be initially investigated within a narrowly
scoped problem context defined by simplifying assumptions on infrastructure,
data, and user interaction. These controlled settings allow for isolated
exploration and validation of core methods. In collaboration with adjacent
research disciplines, such assumptions can then be gradually relaxed. This
step-wise generalization facilitates systematic integration of insights
across domains and enables an assessment of how scientific contributions
generalize to more complex, real-world environments.

From the infrastructure perspective, for instance, research can begin with
centralized computation in a cloud environment, which is sufficient for developing
initial techniques in anomaly quantification and visualization. These methods
can later be extended to the distributed settings by incorporating operator
pipelines at edge devices and hierarchical fragment aggregation in stream
processing architectures, enabling exploration of trade-offs such as latency
and resource constraints.

Similarly, from the data perspective, early work can assume structured,
discrete event data where activities are fully ordered and reliably grouped
per case. This simplifies the development of privacy-aware exchange
mechanisms and process discovery on distributed data sources. Gradually, these assumptions
can be relaxed by incorporating abstraction techniques to handle unstructured
and noisy data.

The user perspective can be approached incrementally, starting from basic
analysis tasks such as process discovery, and progressing to more advanced
scenarios like conformance checking and performance evaluation. Early
research can focus on static data representations; over time, dynamic aspects
such as privacy-aware aggregation and integration with distributed stream
processing and visualization techniques can be introduced.

Advancing this research agenda calls for a principled methodological approach.
Future work on process mining over distributed data sources should build on
algorithm engineering as a foundational framework. In this context, an
algorithm is understood as a well-defined sequence of computational steps
that transforms some input into some output~\cite
{cormen2009introduction,aho1974design}, while engineering refers to the
design and operation of artifacts that meet a recognized need~\cite
{staples2014-ontology}. Accordingly, algorithm engineering is the discipline
that focuses on the design, analysis, implementation, tuning, debugging, and
experimental evaluation of algorithms~\cite
{demetrescu2004algorithm,sanders2009algorithm}. Research in this area should
combine formal analysis with empirical studies, including simulation,
benchmarking, and user-centered evaluation, to systematically explore
trade-offs and validate solutions under realistic constraints.

In summary, the research fields involving process mining on distributed data sources are best
understood as spanning a cross-product of structural shifts and system
dimensions. Rather than addressing each research field in isolation, future
research must pursue integrated solutions that are infrastructure-efficient,
data-sensitive, and user-centric—anchored in the broader vision of
continuous, decentralized process intelligence.

\section{Conclusion}\label{sec:5}
We introduced the concept of process mining on distributed data sources, which enables decentralized, online process mining that processes data as close to the data sources as possible, allowing for timely and continuous analysis. We introduced multiple possible research fields and discussed how interdisciplinary collaboration is required to address the shifts and areas that necessitate awareness when researching process mining on distributed data sources. We observed a shift from centralized, retrospective, event-based mining to distributed, real-time processing of sensor data. Additionally, we observed that the distributed infrastructure, the fine-grained and imprecise sensor data, and the visualizations on the user interfaces require attention. We observed use cases in the domains of logistics, healthcare, and smart manufacturing, which require process mining on distributed data sources to enable real-time process mining on a fragmented, heterogeneous, and dynamic data landscape.
\section*{Acknowledgments}
This work has been funded by the Deutsche Forschungsgemeinschaft (DFG, German Research
Foundation) – FOR 5495.
\bibliographystyle{elsarticle-num}
\pagebreak
\bibliography{bibliography}



\end{document}